\documentclass[aps, pre, amsmath, amssymb, superscriptaddress, a4paper, floatfix, showkeys, 10pt]{revtex4-2}

\usepackage{graphicx}% Include figure files
\usepackage{dcolumn}% Align table columns on decimal point
\usepackage{bm}% bold math
\usepackage{comment}
\usepackage{siunitx}
\usepackage{amsmath}
\usepackage[version=4]{mhchem}
\usepackage{xcolor}
\definecolor{rose}{HTML}{F7879A}

\begin{document}

%\preprint{APS/123-QED}

\title{Shock compression-based equation of state for perfluorohexane}% Force line breaks with \\
%\thanks{A footnote to the article title}%

\author{Anunay Prasanna}
 %\altaffiliation[Also at ]{Physics Department, XYZ University.}%Lines break automatically or can be forced with \\
 \email{aanunay@ethz.ch}
\author{Guillaume T. Bokman}%
\author{Samuele Fiorini}
\author{Armand Sieber}

\affiliation{%
 Institute of Fluid Dynamics, D-MAVT, ETH Z\"urich, Switzerland
 %\textbackslash\textbackslash
}%

%\collaboration{MUSO Collaboration}%\noaffiliation

\author{Bratislav Luki\'c}
 %\homepage{http://www.Second.institution.edu/~Charlie.Author}
%\affiliation{
% Second institution and/or address\\
% This line break forced% with \\
%}%
%\affiliation{
% Third institution, the second for Charlie Author
%}%
\author{Daniel Foster}
\affiliation{%
 European Synchrotron Radiation Facility, Grenoble, France\\
 %This line break forced with \textbackslash\textbackslash
}%

\author{Outi Supponen}
\affiliation{%
 Institute of Fluid Dynamics, D-MAVT, ETH Z\"urich, Switzerland
 %\textbackslash\textbackslash
}%

%\collaboration{CLEO Collaboration}%\noaffiliation

\date{\today}% It is always \today, today,
             %  but any date may be explicitly specified

\begin{abstract}
Perfluorohexane is a biocompatible material that serves as a liquid core for acoustically-responsive agents in biomedical applications.
%However, the development of these agents is significantly affected by the complexity and limited data needed to establish thermodynamic relations, making it challenging to perform numerical simulations to understand the phenomena governing their behavior in these applications. 
Despite its relatively widespread usage, there is a lack of experimental data determining its thermodynamic properties.
This challenges numerical simulations to predict the acoustic response of agents developed using this material.
In this study, we employ the well-established method of shock compression of materials at relatively high pressures (100--400~\si{\mega\pascal}) to estimate a kinematic equation of state for perfluorohexane.
We use multi-objective optimization to obtain the Noble-Abel Stiffened-Gas equation of state, which is suitable for hydrodynamic numerical simulations. % enabling us to complement our experiments with simulations.
We then apply the extrapolated equation of state to simulate shock-wave propagation within a perfluorohexane droplet showing excellent agreement with equivalent experiments.
%This validates the equation of state and promotes the use of numerical simulations as a valuable tool for understanding the complex acoustic interactions involved in these biomedical agents, ultimately facilitating their translation for clinical purposes. 
This promotes the use of numerical simulations as a valuable tool for understanding the complex acoustic interactions involved in these biomedical agents, ultimately facilitating their translation for clinical purposes.
%\begin{description}
%\item[Usage]
%Secondary publications and information retrieval purposes.
%\item[Structure]
%You may use the \texttt{description} environment to structure your abstract;
%use the optional argument of the \verb+\item+ command to give the category of each item. 
%\end{description}

\end{abstract}

%Use showkeys class option if keyword
                              %display desired
\maketitle

%\tableofcontents
 
\section{\label{sec:level1}Introduction}%\protect%\\ The line
%break was forced \lowercase{via} \textbackslash\textbackslash}

Perfluorocarbons have recently gained traction as building blocks for acoustically-responsive agents in various biomedical applications~\cite{Shakya2024Ultrasound-responsiveDelivery}. 
Perfluorohexane (PFH, \ce{C6F14}), a heavier fluorocarbon, has long been investigated for its potential oxygen-carrying capacity~\cite{Lowe2001FluorinatedCarriers} and as phase-change agents in the form of sub-micron droplets, which upon ultrasound activation are converted into microbubbles~\cite{Picheth2017EchogenicityNanosystems}.
This process is typically called acoustic droplet vaporization (ADV)~\cite{kripfgans2000} and is leveraged in several theranostic applications, such as contrast-enhanced imaging, sonodynamic therapy, and targeted drug delivery~\cite{Diaz-Lopez2010Liquid19F-MRI,Namen2021RepeatedImaging,Zeng2020Perfluorohexane-LoadedTherapy,Xiao2024AcousticallyDelivery}.
Compared to the more established microbubbles, nanodroplets present higher circulation times \emph{in vivo} enabling theranostic processes to be implemented for a longer duration~\cite{borden2020}. 
Furthermore, their small size allows them to extravasate into the vascular space due to the Enhanced Permeability and Retention (EPR) effect and to interact directly with cancers and tumor cells, provided that EPR occurs during treatment~\cite{Maeda2015TowardHeterogeneity}.  
However, ADV of droplets based on heavier perfluorocarbons requires high negative pressures for activation, which makes the process susceptible to off-target cavitation bioeffects and, therefore, unsafe for clinical use~\cite{Shakya2024Ultrasound-responsiveDelivery}.

The mechanism of ADV initiation has been attributed to the acoustic impedance mismatch between perfluorocarbons and their surrounding medium (water and blood for example) leading to the focusing of the acoustic wave within the droplet~\cite{shpak2014, Fiorini2024PositiveVaporization}. 
Recently, \citet{Fiorini2024PositiveVaporization} have further shown that focusing of the compression phase of the incoming acoustic wave can generate negative pressures in the droplet bulk.   
This suggests that waves with high positive pressures and minimal tensile excitation such as therapeutic shock waves could be employed as acoustic sources to initiate phase change of droplets.
%The advantage of such shock waves is that they can have a very short duration of high pressure and a weak tensile phase.
Provided that they propagate in the human body with minimal interference and reflections, such shock waves can reduce undesirable bioeffects. 
High-amplitude shock waves, with peak pressures up to 100~\si{\mega\pascal}, have been clinically approved for therapeutic applications such as lithotripsy to treat kidney stones~\cite{loske2017medical}, while shock waves with peak pressures closer to 10~\si{\mega\pascal} are being explored for other delicate purposes such as stem cell and neuron activation~\cite{maier2003substance,beisteiner2020transcranial}, microcirculation stimulation~\cite{kisch2016repetitive} and targeted drug delivery using nanobubbles~\cite{marano2016doxorubicin, marano2017combining}.
Although there has been preclinical research promoting shock wave interactions with nano- and microbubbles, further developments are hindered by their interactions being difficult to analyze due to their small spatiotemporal scales.
This becomes even more complex in shock-induced droplet vaporization, as the physical mechanisms that trigger shock-droplet interactions are not yet fully understood. 
This lack of understanding still poses challenges for clinical translation.
%This is a crucial step to optimize the acoustic driving for nanodroplets and therefore, the clinical translation of employing shock-driven nanodroplets.

Hydrodynamic, multiphase numerical simulations are a powerful tool that can contribute to building a better understanding of shock-droplet interaction and therefore, optimize the acoustic driving necessary to induce phase change within the drop.
However, the modeling requires thermodynamically valid constitutive relations for the material (PFH), \emph{i.e.}, an equation of state (EOS) for closure, which are difficult to calibrate~\cite{Menikoff1988TheMaterials}.
Existing thermodynamic equations of state for PFH are highly uncertain or have limited validity due to the scarcity of experimental data. 
\citet{Gao2021EquationsN-Perfluorohexane} have reported Helmholtz energy equations of state for PFH that are valid for a large range of pressures and temperatures.
However, to ensure the numerical stability of simulations utilizing high-pressure shock waves, leading to pressures larger than standard critical pressures of the material (1.741~\si{\mega\pascal} for PFH~\cite{Gao2021EquationsN-Perfluorohexane}), a more accurate and well-calibrated EOS tailored specifically for such phenomena needs to be obtained~\cite{Cocchi1996TreatmentGodunov-type}.
%Considering that here we utilize high-pressure shock waves, activated at values much higher than the standard critical pressure of PFH , a more accurate and well-calibrated EOS, tailored specifically for such phenomena needs to be obtained, especially to ensure the numerical stability of hydrodynamic simulations~\cite{Cocchi1996TreatmentGodunov-type}.

Shock compression-based experiments, which rely on inducing planar shock fronts in media to derive their EOS, are a well-established method, yet considerably more challenging for liquids than solids~\cite{Rice1957EquationKilobars, Barker1970Shock-waveSapphire, Nellis1980ShockKbar, Mitchell1981EquationRange}.
The Rankine-Hugoniot jump relations between the fluid state before and after the passage of a shock wave are used to establish a kinematic EOS as a base to obtain the desired EOS for numerical simulations \citep{Rankine1997, HugoniotH.Partie}. 
The Rankine-Hugoniot equation of state defines thermodynamic state variables as follows:
\begin{equation}
\label{eq:RH_definition}
\begin{split}
        P - P_\mathrm{0} &= \rho_\mathrm{0}(u_\mathrm{s} - u_\mathrm{0})(u_\mathrm{p} - u_\mathrm{0})\\
    v &= v_\mathrm{0}\left[1 - \frac{u_\mathrm{p} - u_\mathrm{0}}{u_\mathrm{s} - u_\mathrm{0}}\right]\\
    e - e_\mathrm{0} &= \frac{1}{2}(P + P_\mathrm{0})(v_\mathrm{0} - v),
\end{split}
\end{equation}
where $P$ is the pressure, $v$ is the specific volume, $e$ is the internal energy, $\rho$ is the material density, $u_\mathrm{s}$ is the shock wave velocity, and $u_\mathrm{p}$ is the particle velocity in the material behind the shock front. The subscript 0 in Eq.~\eqref{eq:RH_definition} refers to the initial state of the material. 
The EOS employed in numerical simulations governed by the Euler equations must be convex and hyperbolic in order to prevent the sound speed from becoming negative in unstable flows and for the simulation to be scientifically accurate~\cite{Menikoff1988TheMaterials, Saurel2008ModellingFlows}. 
%To use an EoS for closure in numerical simulations governed by the Euler equations, it has to be convex and hyperbolic~\cite{Menikoff1988TheMaterials, Saurel2008ModellingFlows}. 
%This requirement is mainly to stabilize the simulation scientifically for metastable flows and to prevent the sound speed from becoming negative in the unstable region~\cite{Saurel2008ModellingFlows}.
Such an EOS must therefore be developed from experimental data. 
Multi-objective optimization algorithms have recently been used to obtain EOS fits that can capture phase transitions and critical phenomena~\cite{Cox2015FittingIntelligence,Myint2021MinimizationOptimization,Robert2022UsingExample}.
The method has proven adequate to accurately represent material properties.

In this work, we perform planar impact experiments using a mesoscale gas-launcher to first determine the $u_\mathrm{s} - u_\mathrm{p}$ relation for PFH from which we utilize a multi-objective optimization program to derive the Noble-Abel Stiffened-Gas (NASG) EOS~\cite{LeMetayer2016TheState}, which is suitable for hydrodynamic simulations.
%To our knowledge, this is the first work to report such data. 
We qualitatively verify the generated EOS by performing numerical simulations corresponding to the impact experiments as well as a PFH droplet interacting with a laser-induced shock wave, and compare the propagation of the shock front in the simulations with the experiments.
This work establishes the framework to determine other EOS and thermodynamic properties for PFH and opens up prominent avenues for simulation-based study of reagents and nanodroplets using PFH.  

\section{Materials and methods}

\begin{figure*}
\centering
\includegraphics[width=\linewidth]{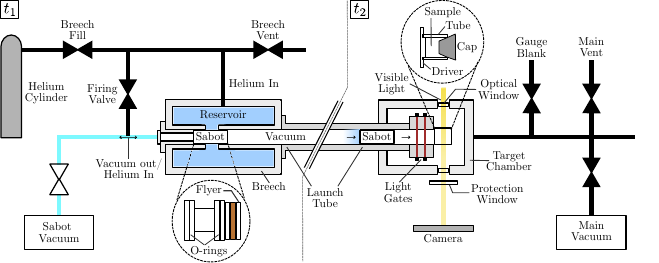}
\caption{\label{fig_1}Schematic of the experimental setup: At time, $t_\mathrm{1}$, the sabot is located in the breech and held by vacuum. Upon filling the breech with Helium and firing (at time, $t_\mathrm{2}$), the sabot is shot forward with velocity $u_\mathrm{f}$ and impacts the container, which is located inside the target chamber, also maintained under vacuum. The light gates are separated by a known distance and enable precise calculation of the impact speed as the sabot passes through them. }
\end{figure*}

%\subsection{Sample preparation}

\subsection{Experimental setup}

\subsubsection{Impact experiments}

The sample container and the sabot for the launcher represented in Fig.~\ref{fig_1} are machined from polymethylmethacrylate (PMMA) and polycarbonate (PC), in accordance with previous experimental work~\cite{Rutherford2017ProbingMesoscale}.
%They are both highlighted in Fig.~\ref{fig_1}.
The container consists of a cylindrical PMMA tube with an outer diameter of 20~\si{\milli\meter} and thickness of 2~\si{\milli\meter}.
The tube is glued to a 1.7~\si{\milli\meter} thick driver plate, which presents a 1.3~\si{\milli\meter} long, 16~\si{\milli\meter} thick extrusion to increase the contact surface between the tube and the plate.
%The driver plate is 1.7~\si{\milli\meter} thick and has six through-holes used to fix the sample container onto a polyamide stand-off ring.
The end cap is 3D-printed with UV-curable resin, has a trapezoidal section, and is also glued to the cylinder.
%The driver plate, the container, and the end cap are glued together using epoxy.
The container is filled with PFH  (Fluoromed, $\rho_\mathrm{PFH} = 1680.1\;\si{\kilo\gram\per\meter\cubed}$) using a through hole in the end cap, which is sealed using epoxy. 
%This assembly is mounted onto a stand-off ring manufactured from polyamide, which in turn is mounted with the help of six M3 screws using the through-holes on the driver plate and tapped holes on the stand-off ring.

The sabot is machined in a dog-bone shape from PC and has a maximum diameter of 25~\si{\milli\meter}. 
Two O-rings (NBR O-ring 18$\times$2.5~\si{\milli\meter}) are wrung around both ends of the sabot to hold it in place under vacuum. 
A PMMA flyer, attached to a copper plate and an additional PMMA piece to increase the sabot's mass and lower its impact speed, is glued to the top of the sabot. 
The PMMA flyer is chosen to match the impedance of the impacting section to that of the container, simplifying the calculation of the particle velocities during post-processing.

The experiments are carried out using a mesoscale gas-launcher used in previous studies at the European Synchrotron Radiation Facility's ID19 beamline~\cite{Rack2024DynamicESRF}.
A schematic showing the single-stage mode of the launcher, powerful enough to achieve the desired impact speeds of the present study, is depicted in Fig.~\ref{fig_1}.
The sabot is loaded into the breech, consisting of a reservoir of high-pressure helium gas and maintained under vacuum at time $t_1$.
Adjusting the pressure of the helium gas allows for tuning the impact velocities of the sabot.
The sample is loaded in the target chamber, also maintained under vacuum, and the base of the driver plate is aligned perpendicular to the sabot's propagation path to ensure planar impacts.
The sample is simultaneously aligned perpendicular to the optical path, with high-speed shadowgraphy (Shimadzu HPV-X2 with LAOWA 2.1$\times$ zoom-lens, 2 -- 5 million fps, $t_\mathrm{ex} = 100- 200$~\si{\nano\second}, 0.4--0.5$\times$ magnification) performed to visualize the shock wave propagation in PMMA and PFH.
Optical windows are machined out of transparent PMMA sheets (2~\si{\milli\meter} thick), and a sturdier PMMA protection window (4~\si{\milli\meter} thick) is used to prevent any exploded material from damaging the high-speed camera setup.
At time $t_2$, the sabot is launched down the 25~\si{\milli\meter} bore barrel connecting the breech to the target chamber.
Along its path, the sabot cuts through two light gates separated by a known distance ($d = 12.1$~\si{\milli\meter}). 
This system is based on two laser diodes connected to two photodetectors (Thorlabs, PDA10A2) having their read-out signal recorded on an oscilloscope (LeCroy Waverunner 9020). 
The recorded timings provide an accurate measurement of the impact velocity and trigger the high-speed camera, thereby completing one measurement.  

\subsubsection{Shock-droplet interaction setup}
\label{subsec:shock_exp_setup}

\begin{figure}
    \centering
    \includegraphics[width=0.5\linewidth]{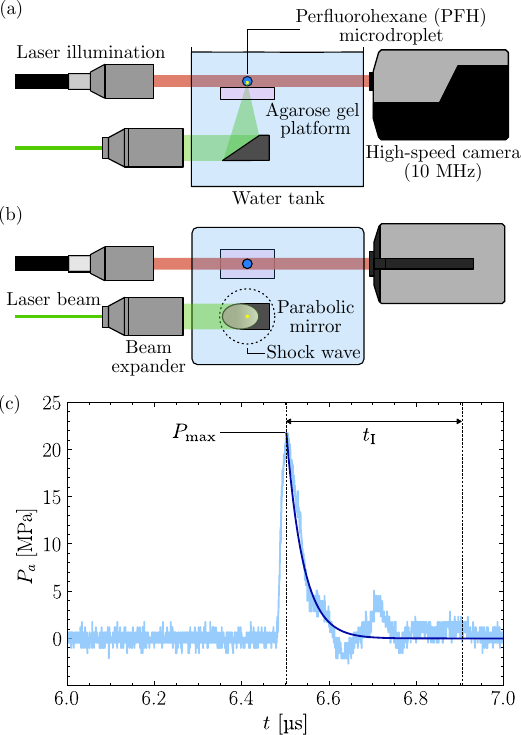}
    \caption{Experimental setup to capture the shock front propagation in a PFH droplet. (a) Side view: A droplet is positioned on a gel platform and imaged with a frame rate of 10 million fps. (b) Top view: The shock wave is generated by laser-induced breakdown at the focal point of a parabolic mirror. The generated spark is positioned at the same height as the droplet and arranged so that the propagating shock front hits the droplet at a position perpendicular to the illumination path. (c) Experimental pressure profile of the shock wave at the droplet's location (light blue line) superimposed by the idealized waveform (dark blue line) computed using the Friedlander equation with $P_\mathrm{max} = 21.7$~\si{\mega\pascal}, $t_I=428$~\si{\nano\second}, and $b = 10$.}
    \label{fig_2}
\end{figure}

The experimental setup employed for visualizing the shock-droplet interaction is depicted in Fig.~\ref{fig_2}.
PFH droplets with radius $R=600-1000$~\si{\micro\meter} are placed on a 2\% w/v agarose gel plate and immersed in a water tank.  
Shock waves are generated, approximately 5~\si{\milli\meter} away from the droplet location, by optical breakdown of water using a frequency-doubled, pulsed Nd:YAG laser (Lumibird Quantel Q-smart, 6~\si{\nano\second}, 532~\si{\nano\meter}). The beam is first expanded using a $10\times$ beam expander (52-71-10X-532/1064, Special Optics) and then focused by a 90\si{\degree} parabolic mirror (Edmund Optics, $f=50$~\si{\milli\meter}) to generate a spherically propagating shock wave.
The shock profile is measured $d = 4$~\si{\centi\meter} away from the focus using a 75~\si{\micro\meter} needle hydrophone (Precision Acoustics, NH0075) and then extrapolated to obtain the peak pressure at the droplet location (see \citet{Bokman2023ScalingWave} for full details).
The shock wave is spherically propagating and shows an exponentially decaying profile as depicted in Fig.~\ref{fig_2}(c).
The Friedlander equation is used to approximate an idealized shock profile as $P_\mathrm{a}(t) = P_\mathrm{max}(1 - t/t_\mathrm{I})\exp{(-bt/t_\mathrm{I})}$, where $P_\mathrm{max}$ is the peak pressure, $t_\mathrm{I}$ is the total duration of the shock wave, and $b$ is a fitting parameter obtained from the hydrophone measurements.
Such a curve is superimposed on top of the experimental measurements with $P_\mathrm{max}=21.7$~\si{\mega\pascal}, $t_\mathrm{I}=428$~\si{\nano\second}, and $b=10$ in Fig.~\ref{fig_2}(c). 
High-speed shadowgraphy is performed to visualize the shock wave within the droplet. A pulsed red laser (645 \si{\nano\meter}) illumination system (Cavitar, Cavilux Smart UHS) with a pulse duration time of $t_\mathrm{ex} = 10$~\si{\nano\second} is used to obtain images of the shock front propagation. The video is recorded at 10 million fps by a high-speed camera (Shimadzu HPV-X2). 
The laser illumination and the high-speed camera are synchronized and both are triggered by the Nd:YAG laser through a delay generator (Stanford Research Systems, DG645).

\subsection{Numerical simulations}
\label{subsec:num_sim}

\begin{figure}[htbp]
    \centering
    \includegraphics[width=0.5\linewidth]{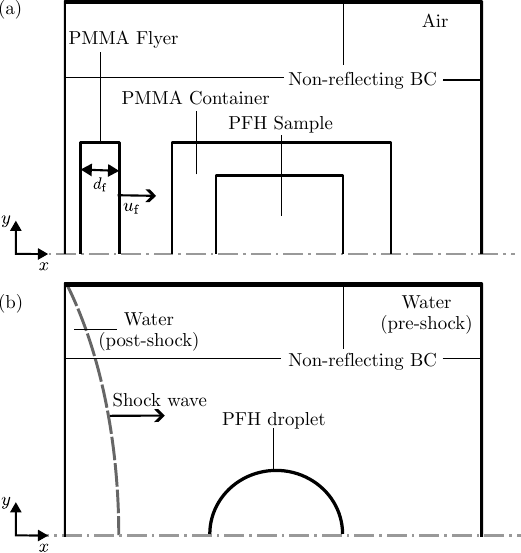}
\caption{\label{fig_3}Schematic depicting the initial numerical setup (a)~Impact experiment: A PMMA flyer of thickness, $d_\mathrm{f}$ with impact velocity $u_\mathrm{f}$ is launched towards a PMMA container loaded with liquid PFH. (b) Shock-droplet interaction experiment: A spherically propagating shock wave, propagating towards the right, is initiated on the left side of the domain and interacts with a circular PFH droplet. All simulations are 2D axisymmetric.}
\end{figure}

Hydrodynamic simulations are performed using ECOGEN, a thermodynamically well-posed, multiphase, compressible, diffuse interface method~\cite{Schmidmayer2020ECOGEN:}. ECOGEN has been validated for several configurations including interactions of bubbles~\cite{trummler2020near} and droplets~\cite{Dorschner2020OnAerobreakup, Bokman2024Impulse-drivenDrops} with shock waves. 
The governing equations are the modified Euler equations with relaxation procedures as described in \citet{Kapila2001Two-phaseEquations} and \citet{RichardSaurel2009SimpleMixtures} and are given as: 

\begin{equation}
    \begin{split}
        \frac{\partial \alpha_j}{\partial t} + \bm{u}\cdot{\nabla \alpha_j} &= \delta p_j\\
        \frac{\partial \alpha_j \rho_j}{\partial t} + \nabla \cdot (\alpha_j \rho_j \bm{u}) &= 0\\
        \frac{\partial \rho \bm{u}}{\partial t} + \nabla \cdot (\rho \bm{u} \otimes \bm{u} + P\bm{I}) &= 0\\
        \frac{\partial \rho E}{\partial t} + \nabla \cdot [(\rho E + P) \bm{u}] &= 0\\
%        \frac{\partial f}{\partial t} + \bm{u} \cdot \nabla f &= 0
    \end{split}
    \label{eq:ECOGEN}
\end{equation}
where the subscript $j = 1,2,\cdots,N$  corresponds to the $N$ phases in the simulation.
The volume fraction and density of each phase are denoted by $\alpha_j$ and $\rho_j$ respectively, and $\delta p_j$ is the pressure-relaxation term for each phase. 
The mixture velocity and pressure are represented by $\bm{u}$ and $P$.
The mixture density is defined as, $\rho = \sum_j \alpha_j \rho_j$.
The total energy is denoted by $E$, and is defined as $E = e + \frac{1}{2}||\bm{u}||^2$, where $e = \sum_j \alpha_j \rho_j e_j$ is the mixture internal energy.
%$f$ is a color function.
%Capillary effects, although negligible are accounted for in the stress tensor $\omega$ defined as
%The first equation of the set signifies the advection of the interface between the two fluids, with the term on the right-hand side ($K\nabla\cdot\bm{u}$) accounting for the acoustic behavior difference between the different phases, with $K$ defined as
%\begin{equation}
%    \Omega = -\sigma \left(||\nabla f||I - \frac{\nabla f \times \nabla f}{||\nabla f||}\right)
%    \label{eq:omega_ECOGEN}
%\end{equation}
%where $c_j$ is the speed of sound of phase $j$ under the given conditions.
Constitutive EOS for the different materials are employed to relate the speed of sound to the pressure in the fluid.
The ideal gas law is used as an EOS for air
\begin{equation}
    \label{eq:ig_air}
    P_\mathrm{air} = \rho_\mathrm{air}(\kappa_\mathrm{air} - 1)(e_\mathrm{air} - e_\mathrm{air,ref})
\end{equation}
where $\kappa_\mathrm{air} = 1.4$ and $e_\mathrm{air,ref} = 0$~\si{\joule\per\kilo\gram} are model parameters~\cite{LeMetayer2004ElaborationDiphasiques} and $\rho_\mathrm{air} = 1.2$~\si{\kilo\gram\per\meter\cubed} is the density of air.
Water is modeled using the stiffened gas EOS
\begin{equation}
    \label{eq:sg_water}
    P_\mathrm{w} = \rho_\mathrm{w}(\kappa_\mathrm{w} - 1)(e_\mathrm{w} - e_\mathrm{w,ref}) - \kappa_\mathrm{w}\Pi_\infty
\end{equation}
where $\Pi_\infty$ is another model parameter, with $\kappa_\mathrm{w} = 2.955$, $e_\mathrm{w,ref} = 0$~\si{\joule\per\kilo\gram}, $\Pi_\infty = 7.22\times10^{8}$~\si{\pascal} and $\rho_{w} = 996$~\si{\kilo\gram\per\meter\cubed} is the density of water.   
Both of these EOSes are well-validated for ECOGEN through several case studies~\cite{Schmidmayer2020ECOGEN:,Dorschner2020OnAerobreakup,trummler2020near,Bokman2024Impulse-drivenDrops}. 
For PMMA and PFH, the Noble-Abel Stiffened-Gas (NASG) EOS is employed \citep{LeMetayer2016TheState}.
\begin{eqnarray}
\label{eq:NASG_1}
    e(P,v) = \frac{P + \gamma P_\infty}{\gamma - 1}(v-b) + q\\
\label{eq:NASG_2}    
    v(P,T) = \frac{(\gamma - 1)C_\mathrm{v}T}{P + P_\infty} + b\\
\label{eq:NASG_3}    
    h(P,T) = C_\mathrm{p}T + bP + q
\end{eqnarray}
where $e,P,v,T,h$ are the internal energy, pressure, specific volume, temperature, and enthalpy of the material respectively.
The other variables are the specific heat at constant pressure and volume, $C_\mathrm{p}, \; C_\mathrm{v}$, and their ratio $\gamma$.
The specific volume parameter, $b$ accounts for the repulsive effects between the molecules of the material, while $P_\infty$ is a stiffness parameter to indicate that materials can withstand a degree of tension without changing phase~\cite{Menikoff1988TheMaterials}. 
The NASG EOS also requires the definition of a reference entropy, $q'$, which we set equal to zero following the convention for liquids~\cite{LeMetayer2016TheState}.
From Eq.~\eqref{eq:NASG_1} and the thermodynamic definition of the speed of sound in the material, we get 
\begin{equation}
\label{eq:NASG_SoundSpeed}
    c^2 = \left(\frac{\partial p}{\partial \rho}\right)_s = \frac{\gamma(P + P_\infty)}{\rho(1 - b \rho)} 
\end{equation}
All the above EOSes are fitted from $u_\mathrm{s}-u_\mathrm{p}$ relations, either available from literature (air, water~\cite{LeMetayer2016TheState} and PMMA \citep{Jordan2016ShockPolymethylmethacrylate}) or from the experimental data collected within the present work (PFH).
The system of equations described above is solved numerically using a splitting procedure with explicit time-integration schemes, which is a standard procedure for such equations. Further details about the numerical scheme employed are fully described elsewhere \cite{Schmidmayer2020ECOGEN:,Bokman2024Impulse-drivenDrops}.

In order to verify the performance of the EOS obtained herein, two-dimensional, axisymmetric numerical simulations of the impact experiments are performed.
A schematic depicting the domain is provided in Fig.~\ref{fig_3}(a).
A PMMA projectile simulating the flyer with thickness 4~\si{\milli\meter} and the same initial velocity as measured in the experiments is launched into a cylindrical PMMA tube (ID = 16~\si{\milli\meter}, OD = 20~\si{\milli\meter}, $\rho_\mathrm{PMMA} = 1185$~\si{\kilo\gram\per\meter\cubed}, $P_1 = 101.3$~\si{\kilo\pascal}) containing liquid PFH ($\rho_\mathrm{PFH} = 1680.1$~\si{\kilo\gram\per\meter\cubed}, $P_\mathrm{0,PFH} = 101.3$~\si{\kilo\pascal}). 
The external medium is simulated as air at ambient conditions ($\rho_\mathrm{air} = 1.2$~\si{\kilo\gram\per\meter\cubed}, $P_3 = 101.3$~\si{\kilo\pascal}).
%Reference properties for this problem are defined at a temperature of $T = 25$~\si{\celsius}.

We further simulate the shock-droplet interaction to check the robustness of the employed EOS with a different case as compared to the experiments used to determine the EOS itself.
This provides the pressure field within the droplet as the shock front propagates in it. 
The PFH droplet ($\rho_\mathrm{PFH} = 1680.1$~\si{\kilo\gram\per\meter\cubed}, $P_{0,\mathrm{PFH}} = 101.3$~\si{\kilo\pascal}) is modeled as a circular discontinuity placed in an infinite medium of water ($\rho_\mathrm{w} = 996$~\si{\kilo\gram\per\meter\cubed}, $P_\mathrm{0,w} = 101.3$~\si{\kilo\pascal}) in a two-dimensional, axisymmetric setting as depicted in Fig.~\ref{fig_3}(b). 
The idealized shock wave pressure profiles (see Fig.~\ref{fig_2}(c)) are initiated as spherically propagating discontinuities in water.
Reference properties for both simulation cases were defined at a temperature of $T=25$~\si{\celsius}.

%Shock waves are visible in shadowgraphs, as they induce variations in the local refractive index, which is proportional to the density gradient~\cite{Gladstone1863}.
By definition, shadowgraphy is the integral of the second derivative of the density along the optical path~\cite{Kleine2006SimultaneousFlows}. 
Computing this numerically would require information about the Gladstone-Dale constant for PFH, which is not available~\cite{Gladstone1863}.
Therefore, we resort to numerical Schlieren to capture the propagation of the shock front in different media, since the shock wave is the source of the largest density gradient within our field of view.
This is evaluated using the equation given below
\begin{equation}
\label{eq:NS}
    S_\mathrm{n} = 1-\exp{\left(\frac{||\nabla\rho||} {||\nabla\rho||_\mathrm{max}}\right)}
\end{equation}
where $||\nabla \rho||$ is the magnitude of the density gradient and $S_\mathrm{n}$ is the numerical Schlieren field, which is then compared to the locations of experimentally visualized shock fronts.

%Since our driving pressures were relatively low and there were several pressure gradients and complex interactions within our domain, it was difficult to directly compare experimental results to the numerical shadowgraphy results in a meaningful manner.
%The schlieren technique is in general more sensitive to gradients than shadowgraphy techniques, and therefore, isolating the numerical shock waves for comparison was relatively straightforward. 

\subsection{Estimation of particle velocity and kinematic EOS for PFH}
\label{subsec:Imp_Match}

\begin{figure}
\centering
\includegraphics[width=0.5\linewidth]{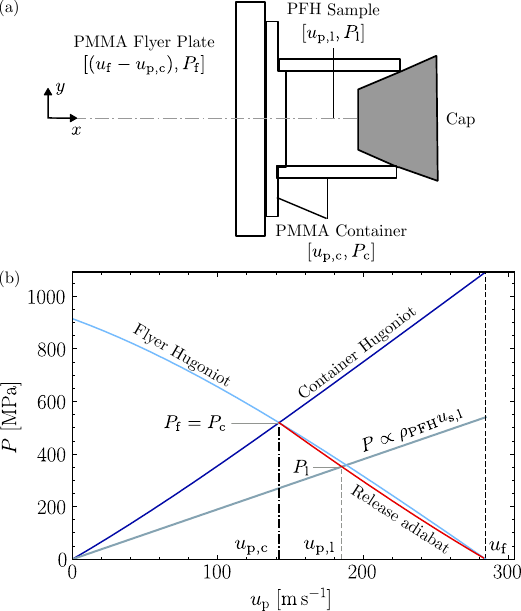}
\caption{\label{fig_4} (a) Schematic depicting the post-shock particle velocity - pressure pairs, $(u_\mathrm{p}, P)$, expected in the flyer, the container and the PFH sample. The particle velocity in PFH is determined by the intersection of the release adiabat, $P_\mathrm{r}$, with a straight line of slope $\rho_\mathrm{PFH}u_\mathrm{s,l}$, where $u_\mathrm{s,l}$ is the shock velocity in the liquid. (b) Graphical depiction of the shock impedance matching technique to determine the expected particle velocity in PFH within the PMMA container for the mesoscale gas launcher experiments.}
\end{figure}

The well-known shock-impedance matching technique~\cite{Menikoff1988TheMaterials, Mitchell1981EquationRange} is used to estimate the particle velocity in PFH and is briefly described below.
Here, the container (with quantities denoted by subscript 'c') and flyer (with quantities denoted by subscript 'f') are designed out of PMMA ($\rho_\mathrm{PMMA} = 1185\;\si{\kilo\gram\per\meter\cubed}$), and the contained liquid (with quantities denoted by subscript 'l') is PFH ($\rho_\mathrm{PFH} = 1680.1\;\si{\kilo\gram\per\meter\cubed}$ at 25~\si{\celsius}).
Since the objects are initially at rest and maintained under vacuum, $u_0$ and $P_0$ are negligible in Eq.~\eqref{eq:RH_definition} and are taken as zero to simplify the calculation.
After impact, all objects are in a post-shock state, with a given particle velocity and pressure as indicated in Fig.~\ref{fig_4}(a).
To estimate the post-shock states of the different objects, first, the Hugoniot pressure of the container is evaluated using Eq.~\eqref{eq:RH_definition} as a function of the particle velocity of the container and its expected shock wave velocity: 
\begin{equation}
\label{eq:Ph_wall}
    P_\mathrm{c} = \rho_\mathrm{PMMA}u_\mathrm{p,c}u_\mathrm{s,c}\;. 
\end{equation}
In general, the shock velocity can be approximated as a polynomial function of the particle velocity. For PMMA, the expression takes the form~\cite{Jordan2016ShockPolymethylmethacrylate}
\begin{equation}
\label{eq:PMMA_SW}
\begin{split}
      u_\mathrm{s} = 6.486u_\mathrm{p}^3 - 7.823u_\mathrm{p}^2 + 3.549u_\mathrm{p} + 2.703, \\
      \text{for} \; u_\mathrm{p} \leq 0.4~\si{\kilo\meter\per\second}  
\end{split}
\end{equation}
where $u_\mathrm{s}$ is the expected shock velocity for a given particle velocity, $u_\mathrm{p}$, in PMMA.
Then, the expected shock velocity in the container, $u_\mathrm{s,c}$, can be evaluated by substituting $u_\mathrm{p} = u_\mathrm{p,c}$ in Eq.~\eqref{eq:PMMA_SW}.
%The validity of this equation for PMMA is justified later. 
At a contact discontinuity, the pressure and particle velocity are considered to be continuous for a planar impact~\cite{Ahrens1993EquationState}.
Therefore, the pressure in the flyer, which has an initial particle velocity of $u_\mathrm{f}$ (equivalent to its launch speed), decreases such that its value is equivalent to the pressure in the container. 
This implies that the particle velocity of the flyer reduces from $u_\mathrm{f}$ to $u_\mathrm{p,c}$, and therefore, the pressure in the flyer can be calculated using Eq.~\eqref{eq:RH_definition} as   
\begin{equation}
\label{eq:Ph_imp}
    P_\mathrm{f} = \rho_\mathrm{PMMA}(u_\mathrm{f} - u_\mathrm{p,c})u_\mathrm{s,f}\;,
\end{equation}
where $u_\mathrm{s,f} = f(u_\mathrm{f} - u_\mathrm{p,c})$ is again estimated using Eq.~\eqref{eq:PMMA_SW}. 
The intersection of the container [Eq.\eqref{eq:Ph_wall}] and flyer Hugoniot [Eq.~\eqref{eq:Ph_imp}] determines the particle velocity in the container, $u_\mathrm{p,c}$, as depicted in Fig.~\ref{fig_4}(b).
It can be shown that $u_\mathrm{p,c} = u_\mathrm{f}/2$.
The release of the container pressure into the liquid stored in it is assumed to follow an isentropic process. 
Since PFH has a lower impedance than PMMA, to ensure the continuity of pressure between the two media, this adiabatic release increases the pressure within the liquid leading to a higher particle velocity than at the container wall.
The particle velocity for such a process can be evaluated from the Riemann integral between the limits of $P_\mathrm{c}$ to the expected pressure in the liquid, $P_\mathrm{l}$ \citep{Menikoff1988TheMaterials}:
\begin{equation}
\label{eq:Riemann_invariant}
    u_\mathrm{r} = u_\mathrm{p,c} + \int_{P_\mathrm{c}}^{P_\mathrm{l}} \frac{dP}{\rho c}\;,
\end{equation}
where $u_\mathrm{r}$ is the particle velocity and $c$ is the sound speed associated with a given pressure along the release adiabat.
Using Eq.~\eqref{eq:RH_definition}, the integral in Eq.~\eqref{eq:Riemann_invariant} can be transformed from the $P-u_\mathrm{p}$ plane into the $u_\mathrm{s}-u_\mathrm{p}$ plane to evaluate $u_\mathrm{r}$ as
\begin{equation}
\label{eq:release_isentrope}
    u_\mathrm{r} = u_\mathrm{p,c} + \int_{u_\mathrm{p,c}}^{u_\mathrm{p,l}} \sqrt{1 - \left(\frac{u_\mathrm{s,c}}{u_\mathrm{p,c}}\frac{\mathrm{d}u_\mathrm{s,c}}{\mathrm{d}u_\mathrm{p,c}}\right)^2} \mathrm{d}u_\mathrm{p}\;,
\end{equation}
where $u_\mathrm{p,c}$ and $u_\mathrm{p,l}$ again represent the particle velocities expected in the container and the PFH liquid. 
Here, the derivative $\mathrm{d}u_\mathrm{s,c}/\mathrm{d}u_\mathrm{p,c}$ is calculated from Eq.~\eqref{eq:PMMA_SW} (since the container is machined from PMMA).
Then, the pressure along the release adiabat is given by
\begin{equation}
\label{eq:rel_pressure}
    P_\mathrm{r} = \rho_\mathrm{PMMA}u_\mathrm{r}u_\mathrm{s,r}\;,
\end{equation}
where $u_\mathrm{s,r} = f(u_\mathrm{r})$ is yet again calculated from Eq.~(\ref{eq:PMMA_SW}). 
Different values of $(u_\mathrm{r},P_\mathrm{r})$ along the isentrope are evaluated by setting different expected values of $u_\mathrm{p,l}$ and calculating the integral given in Eq.~\eqref{eq:release_isentrope}.
The particle velocity in the liquid PFH, $u_\mathrm{p,l}$, is then estimated by the intersection of the release adiabat with a straight line passing through the origin and having a slope of $\rho_\mathrm{PFH}u_\mathrm{s,l}$, as shown in Fig.~\ref{fig_4}(b).

\subsection{Multi-objective optimization for EOS parameter determination}
\label{subsec:MOO}

The estimated kinematic EOS is used to generate the necessary parameters ($b,\;\gamma,\;C_\mathrm{p},\;C_\mathrm{v},\;P_\infty$) for a complete NASG EOS as defined in Sec.~\ref{subsec:num_sim}. 
This is done by reformulating Eqs.~\eqref{eq:NASG_1} and ~\eqref{eq:NASG_2} along with Eq.~\eqref{eq:RH_definition} into cost functions for a multi-objective, least-squares optimization problem.
The formulation is subjected to equality constraints in the form of Eq.~\eqref{eq:NASG_SoundSpeed}, ensuring that the fitted parameters reproduce the correct speed of sound for the material at reference properties extracted from literature~\cite{Jordan2016ShockPolymethylmethacrylate, Marsh2002TemperatureLiquids}. 
An inequality constraint on the material density is added so that the optimization returns physically meaningful results.
Due to the inverse dependence of the sound speed on the density [see Eq.~\eqref{eq:NASG_SoundSpeed}], an absolute equality constraint for both these quantities cannot be enforced in the optimization search.
Therefore, a tolerance of $\pm 5\%$ is provided on the material density.
This tolerance did not affect the numerical simulations, since the evolution of the material density is governed by the continuity and momentum equations.
Without this constraint, a stronger restriction is placed on choosing the lower and upper bounds of the optimization, which is non-trivial for a fluid such as PFH that has limited available literature about its material properties. 
The complete system is expressed as follows: 
\begin{align}
    \begin{split}
        \mathrm{min}\;\; f_1 &= \sum [(e - e_0)_\mathrm{exp} - (e - e_0)_\mathrm{NASG}]^2 \\
        \mathrm{min} \;\; f_2 &= \sum [(v - v_0)_\mathrm{exp} - (v - v_0)_\mathrm{NASG}]^2\\
        \mathrm{s.t.} \;\; P_0 + P_{\infty} &= \left[\left(1 - \frac{C_\mathrm{p} - C_\mathrm{v}}{C_\mathrm{p}}\right)\rho_0c_0^2(1 - b\rho_0)\right]\\
        |\rho_\mathrm{0,ref} - \rho_\mathrm{0,NASG}| &\leq 0.05\rho_\mathrm{0,ref}\\
        C_\mathrm{p}/C_\mathrm{v} - \gamma &= 0\;.   
    \end{split}
    \label{eq:MOO}
\end{align}

The multi-objective optimization problem is solved using the PyMOO module in Python \cite{Blank2020Pymoo:Python}.
A Non-sorted Genetic Algorithm (NSGA-II) is used to find the parameters for the NASG EOS~\cite{Deb2002ANSGA-II}.
To reach the Pareto optimal solution as efficiently as possible, a simulated binary crossover (SBX) with polynomial mutation is applied to improve the quality of the offspring produced at every generation~\cite{Deb2007Self-adaptiveOptimization}. 
The crossover and mutation parameters were set to balance the creation of feasible offspring while exploring the objective space efficiently, without excessive computational cost.  
NSGA-II has been tested on several bi-objective problems and is shown to be highly computationally efficient due to its elitist approach, and therefore, chosen as the most suitable algorithm to solve Eq.~\eqref{eq:MOO}~\cite{Deb2002ANSGA-II, Deb2007Self-adaptiveOptimization}. 

To obtain a suitable termination criterion and to ensure the quality of the obtained optimization results, it is necessary to monitor the performance of the run.
Typically, this involves evaluating certain performance metrics with respect to a reference set (usually the known Pareto front) generation-by-generation. 
However, this requires \emph{a priori} knowledge of the Pareto optimal solution, which is not possible for our optimization problem.
Therefore, the simulation convergence is monitored using running metrics such as the normalized ideal and nadir points, and the normalized, redefined Inverted Generational Distance (IGD) between successive generations \cite{Blank2020AAlgorithms}.  
As a suitable termination condition, we monitor the movement of the objective functions and the above-mentioned parameters per consecutive generations and terminate the optimization when their movement lies below a set tolerance for a given number of successive generations. 
Termination and convergence criteria are discussed further in Sec.~\ref{sec:Results} and Appendix~\ref{app:convergence}.

\section{Results}
\label{sec:Results}

As the sabot is launched and impacts the sample container in the mesoscale gas-launcher experiments, forward- and backward-propagating planar shock waves are produced at the contact discontinuity.
Owing to the higher speed of sound in PMMA as compared to PFH, the forward-propagating wave travels faster in the PMMA container than in the PFH sample.
The backward-propagating wave is further reflected from the attached copper mass on the sabot, while the forward-propagating wave is deflected at the edges of the container, propagating through the epoxy coating used to hold the container together.
The combined effects of the waves promote complex acoustic interactions within the container leading to its cracking.
The cracks expose the sample to vacuum, thereby causing a sudden pressure drop in the liquid that results in cavitation.
Consequently, only the initial forward-propagating shock is employed to estimate the kinematic EOS of both materials and the other complex waves are neglected.
Thirteen shots with impact velocities between 104--280~\si{\meter\per\second} are used to estimate the necessary quantities.
%The section is organized in the order of the results needed to determine the NASG EOS for PMMA and PFH and is concluded by the numerical validation of the determined EOS.

\subsection{Shock-wave propagation in PMMA}
\label{subsec:SW_in_PMMA}

\begin{figure}[htbp]
\centering
\includegraphics[width=0.5\linewidth]{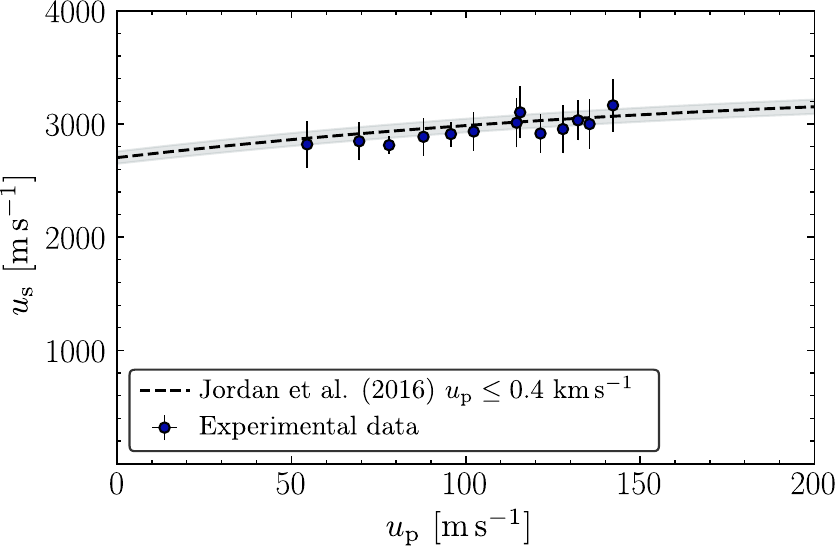}
\caption{\label{fig_5} Experimentally determined shock wave velocity in the PMMA container compared to the EOS obtained from \citet{Jordan2016ShockPolymethylmethacrylate}. Error bars represent the experimental uncertainty associated with the shock velocity measurements. The gray shaded area represents the error expected by \citet{Jordan2016ShockPolymethylmethacrylate} in their results.}
\end{figure}

\begin{figure*}
\centering
\includegraphics[width=\linewidth]{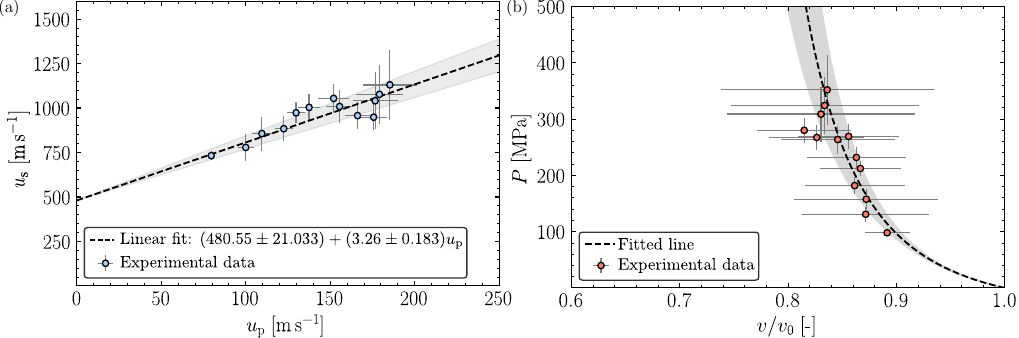}
\caption{\label{fig_6} (a) Kinematic equation of state depicted in terms of the shock velocity, $u_\mathrm{s}$, as a function of the particle velocity, $u_\mathrm{p}$, for PFH. The dashed line plots the linear fit obtained using orthogonal distance regression, with the shaded area indicating the uncertainty associated with it. This uncertainty arises as a direct consequence of considering the experimental errors in both $u_\mathrm{s}$ and $u_\mathrm{p}$ while performing the fitting. (b) The Hugoniot for PFH is plotted in terms of the pressure versus relative volume, with the dashed line showing the expected theoretical curve obtained by evaluating the shock velocity from the linear fit depicted in (a). The gray shaded area again depicts the associated uncertainty similar to (a).}
\end{figure*}

The velocity of the initial shock wave traveling in the PMMA container is measured from the experimental shadowgraph images for different impact velocities (see video in Supplemental Material~\cite{noauthor_notitle_nodate} and Fig.~\ref{fig_7} for sample shadowgraphs) and plotted in Fig.~\ref{fig_5}.
Since our experimental particle velocities are low, the dependency between the shock wave and particle velocity in PMMA is non-linear \cite{Barker1970Shock-waveSapphire, Jordan2016ShockPolymethylmethacrylate}. %Murphy2018NovelLoading, Escauriza2020CollapseLoading}. 
As depicted in Fig.~\ref{fig_5}, our measured shock wave velocities correspond quite well to the cubic polynomial relation connecting the shock and particle velocity in PMMA for $u_\mathrm{p} < 0.4~\si{\kilo\meter\per\second}$ obtained from \citet{Jordan2016ShockPolymethylmethacrylate} and given by Eq.~\eqref{eq:PMMA_SW}. 
The shaded gray area shows the 2\% uncertainty that ~\citet{Jordan2016ShockPolymethylmethacrylate} expect from their results, while the error bars for the experimental values arise from the uncertainty associated with the distance travelled by the shock waves between two frames.
Details for the calculation of our experimental errors are provided in Appendix~\ref{app:error}. 

Fig.~\ref{fig_5} validates Eq.~\eqref{eq:PMMA_SW} as the kinematic EOS of PMMA for our experiments to determine the necessary thermodynamic state variables from the Rankine-Hugoniot set of equations [Eq.~\eqref{eq:RH_definition}].
This is then transposed to the experimentally defined state variables of the minimization functions as described in Eq.~\eqref{eq:MOO} to determine the NASG EOS of PMMA for our numerical simulations.

\subsection{Kinematic EOS of PFH}

\begin{figure*}
    \centering
    \includegraphics[width=\linewidth]{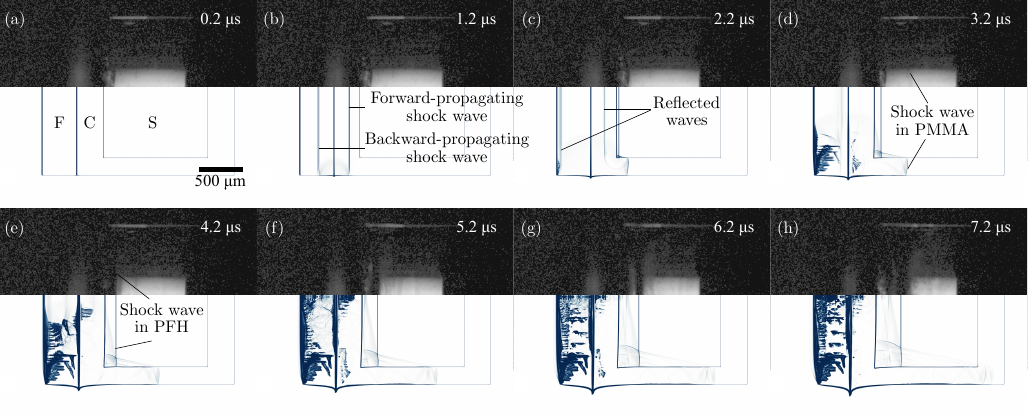}
    \caption{Comparison of the experimental shadowgraphs \emph{(top)} to the numerical Schlieren \emph{(bottom)} for the test case with initial impact velocity of $u_\mathrm{f} = 109$~\si{\meter\per\second}. Here, $t=0$ indicates the instant when the impact occurs. (a) Moment shortly after impact. The PMMA flyer (F), PMMA container (C), and PFH sample (S) are labeled for reference. (b) Upon impact, forward- and backward-propagating shock waves are generated. (c) The interaction of these waves with the container wall and the walls of the flyer induces reflected waves, some of which are visible in the simulation. (d) The shock wave propagating in the PMMA container. (e) The shock wave propagating in PFH becomes clearly visible in the experiment. (f-g) Continued propagation of the waves and multiple spurious reflections due to complex acoustic interactions. Cavitation of the liquid, due to the reflected waves and the tensile tail of the shock wave, leads to a darkened image trailing the shock wave.}
    \label{fig_7}
\end{figure*}

Similar to the shock wave propagation in PMMA, the shock velocity in PFH ($u_\mathrm{s,PFH}$) is estimated from the high-speed images obtained experimentally (see Fig.~\ref{fig_7} for sample shadowgraphs).
This process is repeated for all the different impact velocities recorded experimentally.
Fig.~\ref{fig_6}(a) plots the dependence of the shock velocity on the particle velocity in PFH. 
The error bars represent the maximum uncertainty associated with the shock velocity and the particle velocity, estimated using a Monte-Carlo error propagation technique as detailed in Appendix~\ref{app:error}. %\cite{Mitchell1981ShockTantalum}.
The shock velocity in PFH varies approximately linearly with the particle velocity, although this becomes unclear for higher particle velocities, where the experimental errors become significant.
As mentioned previously, the PMMA container containing the liquid ruptures, which occurs at earlier times with increasing impact speeds.
There could additionally be vessel stretching before rupture that is rather difficult to identify in the videos.
Furthermore, the crack formation and propagation in the container are enhanced at higher particle velocities, leading to the earlier onset of cavitation within the liquid, which again promotes the breaking of the container and could potentially introduce errors while measuring the shock wave speed.

Orthogonal distance regression~\cite{Boggs1990} is used to obtain a linear fit of the form $u_\mathrm{s} = c_0 + au_\mathrm{p}$, with $c_0$ depicting the estimated base speed of sound. 
This method was preferred to account for the experimental errors in both $u_\mathrm{p,PFH}$ and $u_\mathrm{s,PFH}$, and to avoid biasing the fitting process towards the highly uncertain values at larger particle velocities. 
Therefore, the kinematic EOS for PFH can be given as 
\begin{equation}
\label{eq:kin_EOS}
    u_\mathrm{s,PFH} = (480.55\pm21.033) + (3.26\pm0.183)u_\mathrm{p}.     
\end{equation}
We decided not to constrain the base speed of sound for the fit, although constraining it to reference speeds of sound observed between the temperatures of $T=20-25$~\si{\celsius} did not significantly alter the slope of the line.
The associated experimental errors imply that the obtained kinematic EOS is more accurate for lower particle velocities, and therefore at lower Mach values.

Fig.~\ref{fig_6}(b) depicts the $P-v$ Hugoniot up to the pressure of 500~\si{\mega\pascal}, with the dashed line indicating the expected Hugoniot evaluated from combining Eq.~\eqref{eq:RH_definition} with Eq.~\eqref{eq:kin_EOS}.
Although the maximum error in the relative volume reaches 12\%, the overall fit remains satisfactory. 
The thermodynamic state variables of PFH can now be evaluated using Eq.~\eqref{eq:RH_definition} in conjunction with Eq.~(\ref{eq:kin_EOS}) transposed to Eq.~(\ref{eq:MOO}), and used to fit the NASG EOS for PFH.

\subsection{Fitting parameters for NASG EOS}
\label{subsec:MOO_results}

Once the $u_\mathrm{s} - u_\mathrm{p}$ relations for both materials are determined, we can formulate our optimization cost functions, $f_1$ and $f_2$ defined in Eq.~\eqref{eq:MOO}.
This is done by selecting the appropriate reference state of the materials in order to define the constraints of the problem. 
First, we set the reference temperature at which the respective experiments are performed as $T=25$~\si{\celsius}. 
Then, we determine the speed of sound, density, and an approximate range of specific heats for the materials~\cite{pmmaNIST, Gao2021EquationsN-Perfluorohexane}. 
Selecting a slightly higher or lower reference temperature ($\pm2$~\si{\celsius}) did not alter the results, but some cases required the speed of sound to be adjusted at the corresponding temperature, as discussed later.
Since multiple combinations of the equation parameters can be possible solutions for the NASG problem, we set the upper and lower bounds of our fitted parameters using the information available above.
This constrains the optimization between known bounds and prevents the search algorithm from exploding.
The specific heats ($C_\mathrm{p},\;C_\mathrm{v}$) are given a small range in between the experimentally measured values obtained from literature~\cite{pmmaNIST,Gao2021EquationsN-Perfluorohexane}, while the other parameters ($b,\;\gamma,\;P_\infty$) are given ranges similar to values expected for other materials~\cite{LeMetayer2004ElaborationDiphasiques,Saurel2008ModellingFlows,LeMetayer2016TheState}.

\begin{comment}
\begin{figure}[t]
\centering
\includegraphics[width=\linewidth]{Images/figure7.pdf}
\caption{\label{fig_7} Normalized Pareto front of the NSGA-II optimization. The normalized minimization functions are depicted by $\widetilde{f_1}$ and $\widetilde{f_2}$. Equal weightage is given to both functions during the decomposition, which implies that the selected set of parameters represents $\widetilde{f_1}=\widetilde{f_2} = 0.5$, as indicated by the dashed lines. }
\end{figure}
\end{comment}

The optimization is performed using the NSGA-II algorithm with the evolutionary operators selected so as to speed up the computation, as described previously in Sec.~\ref{subsec:MOO}.
A population size of 400 offspring per generation is employed with simulated binary crossover and polynomial mutation operators to enhance the diversity of the feasible children~\cite{Deb2002ANSGA-II}.
The convergence of the optimization program is monitored using the quantities described in Appendix~\ref{app:convergence}. 
%This allows for the efficient selection of operators that promote diverse solutions at a faster pace, which also reduces the computational overhead.
The tolerance for convergence is set as $\epsilon = 0.001 = 0.1\% $ and the simulation is terminated when the average tolerance of the last $\chi = 100$ generations is less than $\epsilon$~\cite{Blank2020AAlgorithms}.
A further decrease in $\epsilon$ or an increase in $\chi$ did not significantly influence the results. 
The optimization is also allowed to run for longer durations ($n_\mathrm{gen} = 50000,100000, 200000$) by ignoring the termination conditions set above to check if there is any significant movement in the objective or design space at later generations. 
This is found to be unimportant, implying that our set termination criteria is satisfactory.

\begin{table}[b]%The best place to locate the table environment is directly after its first reference in text
\caption{\label{tab:table1}%
Values of constants required to define the NASG EOS for PMMA and PFH.   
}
\begin{ruledtabular}
\begin{tabular}{cccc}
\textrm{Constant}&
\textrm{PMMA}&
\textrm{PFH}\\
\colrule
$\gamma$ [-] & 1.416 & 1.477\\
$b$ [\si{\meter\cubed\per\kilo\gram}] & $7.319\times10^{-4}$ & $2.04 \times 10^{-4}$\\
$C_v$ [\si{\joule\per\kilo\gram\per\kelvin}] & 1033.423 &  515.615\\
$C_p$ [\si{\joule\per\kilo\gram\per\kelvin}] & 1463.327 &  761.789\\
$P_{\infty}$ [\si{\pascal}] & $8.1099 \times 10^8$ &  $1.756 \times 10^8$\\
$q$ [\si{\joule\per\kilo\gram}]& $-436.142 \times 10^3$  & $-227.007 \times 10^3$\\
\end{tabular}
\end{ruledtabular}
\end{table}

\color{black}
%Fig.~\ref{fig_7} depicts the normalized Pareto front obtained for PFH using reference properties at 25~\si{\celsius} ($\rho_\mathrm{PFH} = 1680$~\si{\kilo\gram\per\meter\cubed}, $c_0 = 484.975$~\si{\meter\per\second}, $P_0=101.3$~\si{\kilo\pascal}). 
Once the Pareto front has been obtained, suitable parameters for the EOS must be selected. 
We employ the in-built Augmented Scalarized Function (ASF) option in PyMOO~\cite{Blank2020Pymoo:Python}. 
ASF decomposes the multi-objective problem into single objectives and scalarizing functions are derived from a reference point in the objective space.
These scalarizing functions are such that they have their minima located at Pareto points only, thereby allowing the selected point in the objective space to be a Pareto-optimal solution~\cite{Wierzbicki1980TheOptimization}.
Equal weightage is provided to both objective functions and ASF is then applied to obtain the parameters.
%The selected point lies on the Pareto front indicating that the selected parameters denote a global minimum for the optimization problem.
The selected parameters used to define our NASG equations for PMMA and PFH are listed in Table~\ref{tab:table1}, which are then used for the numerical simulations in ECOGEN.

\subsection{Numerical Results}
\label{subsec:num_val}

\color{black}
\subsubsection{Impact experiments}

As a first qualitative verification, \color{black} the impact experiment is simulated with a PMMA flyer launched at a given impact speed $u_\mathrm{f}$ into a PMMA container containing the PFH sample.
%Along with the aforementioned reference properties (see Sec.~\ref{subsec:num_sim}),
The initial reference densities of the materials are evaluated from Eq.~\eqref{eq:NASG_2}, using the constants obtained by the optimization and given in Table~\ref{tab:table1}.
The speed of sound for the different media are defined as follows: $c_\mathrm{0,air} = 343$~\si{\meter\per\second}, $c_\mathrm{0,PMMA} = 2703$~\si{\meter\per\second}~\cite{Jordan2016ShockPolymethylmethacrylate}, $c_\mathrm{0,PFH} = 484.975$~\si{\meter\per\second}. %~\cite{Marsh2002TemperatureLiquids}.
\color{black}
The initial velocity of the container and the sample are set as $u_0 = 0$~\si{\meter\per\second}.
The epoxy resin used as an adhesive to secure the container, along with the attached copper mass of the sabot and the container's end cap, are omitted from our numerical setup.
Including these elements increases the complexity and requires additional EOS for the materials (copper and resins), without significantly influencing the shock wave transmission and propagation into the PFH sample.

\begin{figure*}
    \centering
    \includegraphics[width=0.95\linewidth]{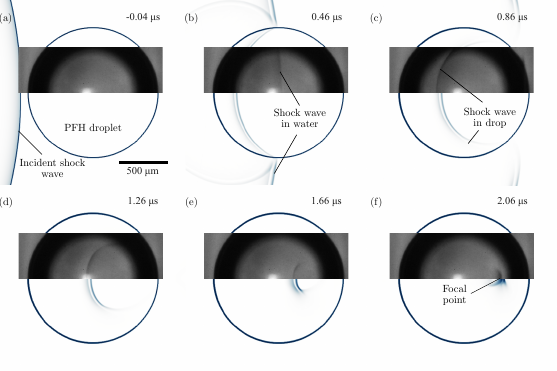}
    \caption{Comparison of the experimental shadowgraphs \emph{(top)} to the numerical Schlieren \emph{(bottom)} for the test case with a laser-induced shock wave in water interacting with a PFH droplet of radius $R = 662$~\si{\micro\meter}. Here, $t=0$ indicates the instant the shock first interacts with the droplet. (a) A decaying, spherically propagating shock is initiated behind the PFH droplet. (b) The shock wave travels faster in water compared to the droplet and is visualized here. (c) The shock shape inside the droplet is clearly visible. The shock propagates further in the droplet (d--e) before focusing and amplifying the pressure at the focal point indicated in (f). }
    \label{fig_8}
\end{figure*}

A qualitative comparison between the experiments and numerical simulations is depicted in Fig.~\ref{fig_7} for an impact speed of $u_\mathrm{f} = 109$~\si{\meter\per\second}.
The results discussed herein are equivalently valid for the other tested impact speeds (109--280~\si{\meter\per\second}).
Snapshot (a) shows the moment shortly after impact, which occurs at time $t = 0$.
The impact generates a forward-propagating shock wave in the PMMA container labeled C in Fig.~\ref{fig_7}(a)] and a backward-propagating shock wave in the PMMA flyer [labeled F in Fig.~\ref{fig_7}(a)] as seen from the numerical Schlieren snapshot in Fig.~\ref{fig_7}(b).
These waves are reflected at the back of the flyer and the container wall as shown in Fig.~\ref{fig_7}(c).
These waves are not clearly visible in the experiments as their magnitude is expected to be lower than the still propagating forward shock wave, which is visualized in the shadowgraphs for the first time in Fig.~\ref{fig_7}(d), and whose location matches well with the numerical result.
This qualitatively verifies \color{black} our fitted NASG equation for PMMA obtained using Eqs.~\eqref{eq:RH_definition}, \eqref{eq:PMMA_SW}, and \eqref{eq:MOO}.
In the numerical Schlieren, the shock wave propagating in PFH is already visible from Fig.~\ref{fig_7}(c), while it is initially obscured by the shadow of the container and the epoxy holding it in the experimental shadowgraphs and only becomes visible in the shadowgraph at a later stage in Fig.~\ref{fig_7}(e).
The correspondence between the experimental and numerical shock wave is reasonable, with the numerical shock wave being slightly slower.
This correspondence is deemed acceptable considering the complexity of the problem, and the minor discrepancies could be attributed to the measured experimental errors and the fact that the EOS obtained for the numerical simulation is a statistical fit.
%The implications of each of these possibilities are further discussed in Sec.~\ref{sec:disc}.

The numerical snapshots from Fig.~\ref{fig_7}(c-h) include large density gradients that continuously grow over time, at the corners of the flyer and the container.
They arise due to the complex interaction of the reflected waves from the contact interfaces in the simulation and are deemed unphysical. 
Complex wave interactions did exist in the experiments and led to several other phenomena such as the cracking of the container, spallation, and the formation of cavitation bubbles within the PFH sample. 
%Indeed, we did notice a few sporadic instances of these during our experiments. 
Nonetheless, the number of frames available to reliably measure the shock wave speed is always sufficient.
The reflected expansion waves are numerically spurious, making the simulations unstable over time.
Modeling the attached copper mass and epoxy resin might help increase the amount of time the simulation remains stable.
However, these interactions coincide with the cracking of the container and the phase change experienced by the sample in the experiments.
The accurate numerical modeling of these phenomena is out of the scope of the study as they do not affect the determination of the EOS for PFH.

\subsubsection{Shock-droplet interaction}

We further employ our NASG EOS for PFH in another test case.
In the shock-droplet interaction experiment, a laser-induced shock wave is generated in water and allowed to propagate into the PFH droplet.
As mentioned previously in Sec.~\ref{subsec:shock_exp_setup}, the waveform of the shock wave is measured experimentally using the hydrophone and is approximated using the Friedlander equation.
%as $p_\mathrm{a}(t) = p_\mathrm{max}(1 - t/t_\mathrm{I})\exp{(-bt/t_\mathrm{I})}$, where $p_\mathrm{max}$ is the peak pressure, $t_\mathrm{I}$ is the total duration of the shock wave, and $b$ is a fitting parameter obtained from the hydrophone measurements~\cite{Bokman2023ScalingWave} 
The results below are given for a shock wave of peak pressure, $P_\mathrm{max} = 21.7$~\si{\mega\pascal}, with $t_\mathrm{I} = 428$~\si{\nano\second} and $b=10$.
% and having a full-width-at-half-maximum (FWHM) of $\delta_a = 39$~\si{\nano\second} at the drop location. The experimental pressure profile and the Friedlander approximation are showed in Fig.PUTFIGURE.
Here, we define the reference densities from Eq.~\eqref{eq:NASG_2}, and the  reference speeds of sound for water and PFH as: $c_\mathrm{0,w} = 1481$~\si{\meter\per\second}~\cite{LeMetayer2004ElaborationDiphasiques}, $c_\mathrm{0,PFH} = 484.975$~\si{\meter\per\second}.

\color{black}
Fig.~\ref{fig_8} qualitatively compares the experimental shadowgraphs to the numerical Schlieren results.
A spherically propagating shock wave (traveling left to right in the images) is initiated behind the droplet and reaches it at $t=0$.
The shock wave, having a faster speed in water than in PFH, is visible in the experiment in Fig.~\ref{fig_8}(b).
The shock wave in PFH is obscured by the dark edge of the droplet (caused by the refraction from the curved interface) in Fig.~\ref{fig_8}(b), but becomes clear in the next snapshot [see Fig.~\ref{fig_8}(c)]. 
The acoustic impedance of water is much higher than PFH and therefore, as the shock wave crosses from water into the droplet, it converges and focuses at the geometrical focus, also known as the focal point of the droplet.
Here, it is expected that the incoming shock pressure is amplified and reaches a maximum value.
Recent studies have shown that acoustic waves, especially the ones with a broadband frequency content, can present sign inversion of the wave while crossing the focal point~ \cite{Fiorini2024PositiveVaporization, Schmidmayer2023}. 
Considering that a shock wave is a sharp, highly broadband wave with a large positive peak pressure, its inversion could lead to large local negative amplification within the droplet.  
In the simulation depicted in Fig.~\ref{fig_8}, we obtain a positive pressure amplification factor of, $|P^{+}_\mathrm{foc}/P_\mathrm{max}| = 4.11$, and a negative pressure amplification of $|P^{-}_\mathrm{foc}/P_\mathrm{max}| = 0.45$ at the focus.
This gives a negative pressure value of $P^{-}_\mathrm{foc} = -9.72~\si{\mega\pascal}$, which could be sufficient to induce droplet vaporization at human body temperature. 
It must be noted that this negative pressure is generated from a purely positive wave crossing the focal point~\cite{Fiorini2024PositiveVaporization}.
\color{black}
The correspondence between the shadowgraphs and the numerical Schlieren is excellent and consistent over multiple shock pressures (15--25~\si{\mega\pascal}) and droplet radii (600--1000~\si{\micro\meter}) considered here.

\section{Discussion}
\label{sec:disc}

In this work, we have obtained the relationship between the shock and particle velocity for PFH by performing planar impact experiments using a single-stage mesoscale gas-launcher.
As mentioned in Sec.~\ref{sec:Results}, the experimental errors associated with the shock velocity measurements increase at higher impact speeds, where the shock wave is captured in fewer frames and where motion blur effects become more notable within the camera exposure time ( $t_\mathrm{ex} = 100- 200~\si{\nano\second}$).
The exposure times employed here are typically longer than what is ideal to capture sharp shock fronts. 
The jitter associated with the expected impact velocity for a given helium pressure in the reservoir of the mesoscale gas-launcher meant that longer exposure times and lower recording speeds are needed to ensure that the shock wave is present in the recorded video.
This implies that the obtained shock wave is relatively thick and introduces variability in the measured shock wave speed.
The widening of the shock front is more apparent for larger impact velocities, which could also result as a consequence of the impact not always being fully planar.
%As mentioned in Sec.~\ref{sec:Results}, the experimental errors associated with the shock velocity measurements are larger with fewer frames capturing the shock wave at higher impact speeds due to increased motion blur within the exposure time.
Additionally, in some cases, the container cracks, causing cavitation within the sample, which implies that the shock wave can only be noticed for short travel distances, further reducing the available frames for measuring the shock wave speed. 
\color{black}For improved measurements of higher shock wave speeds, one could resort to non-optical techniques such as Photon Doppler Velocimetry~\cite{Dolan2020ExtremePDV, Kilic2024TimeMeasurements}, which would also permit using a sturdier container that could sustain such high impact speeds without breaking, thus preventing cavitation in the liquid.
Nevertheless, the experimental errors introduced are considered when fitting the $u_\mathrm{s}-u_\mathrm{p}$ curve for PFH, minimizing the bias from higher impact speeds, and thereby ensuring that the kinematic EOS is valid for lower Mach numbers, as seen from the correspondence between our experiments and simulations for both the impact and the shock-droplet experiments. 

%For the numerical simulations of the impact experiments, a simpler design than the experiments was employed to model the flyer and the container, with the end cap being completely ignored.
%This simplified the complex acoustic interactions involved in the experiment and indeed, made the numerical simulations more stable.
%It allowed for better comprehension of the numerical results and made it easier to compare the numerical and experimental results. 
%Using the complete experimental design would involve more complex modeling of the material at a much higher computational cost, and was deemed unnecessary for the validation.

Aligned with applications of perfluorocarbon droplets for biomedical applications involving acoustic excitation, the EOS has also been employed to simulate a shock wave interacting with a sub-millimetric PFH droplet, and the results compared with ultra-high-speed shadowgraphy. The transmitted shock wave is focused and undergoes a sign inversion as it crosses the focal point, leading to the creation of large negative pressure, which is on the same order of magnitude as the peak pressure of the shock wave. 
This could enable the acoustic vaporization of the droplet, a process shown to have great potential in biomedical applications~\cite{kripfgans2000, shpak2014, Shakya2024Ultrasound-responsiveDelivery}.
\color{black}Here, in the numerical simulations performed, the selected reference speed of sound for PFH is 5\% lower than the expected sound speed of 510.5~\si{\meter\per\second} at 25~\si{\celsius}~\cite{Marsh2002TemperatureLiquids}, which is justified by the excellent match between experiments and numerical simulations.
A change in the reference sound speed of materials by up to $\pm10\%$ is deemed reasonable for low Mach flows, $\mathrm{M}\approx 1+\beta$, with $\beta\ll1$, as noted in other shock-interaction numerical studies~\cite{Cocchi1996TreatmentGodunov-type, Johnsen2008Shock-inducedLithotripsy}.
The correspondence between experiments and numerical simulations confirms that our fitted NASG EOS is valid at low Mach numbers (here $\mathrm{M} \approx 1$).

A comparable NASG EOS could also be fitted by using the derived saturation properties of PFH from the work of \citet{Gao2021EquationsN-Perfluorohexane} following the methodology outlined in \citet{LeMetayer2016TheState}.
The EOS obtained in this manner also requires an adjustment in the reference speed of sound at a given temperature to match the corresponding simulations with the experiments described here.
This EOS works equivalently well for the shock-droplet case when lower input pressures for the shock wave are used as compared to the intensities tested in this study (with $\mathrm{M} \approx 1$).
A good method to check the validity of both EOSes for the shock-droplet interaction would be to compare the pressure or a similar quantity evaluated in the numerical simulation with that in the experiments.
However, measuring such quantities in the interior of a sub-millimetric droplet is non-trivial, making such a comparison challenging and out of the scope of this paper.
It must be noted that the EOS of \textcite{Gao2021EquationsN-Perfluorohexane} is obtained by fitting the data of the material properties of PFH available from several different experiments, all measuring different quantities.
This raises the question of its validity for cases where input pressures larger than the critical pressure of PFH are employed.
While it is challenging to demonstrate significant improvements provided by our EOS, the advantage of our EOS is that it is obtained by shock-based experiments, which are then employed to study other shock-based experiments.
Therefore, while difficult to validate, we expect deviations between the two EOSes at higher Mach numbers, and believe that the EOS derived in this study is more robust for simulations at higher input pressures.
%However, it fails to reproduce experimental results at higher input pressures and higher Mach numbers.
%Therefore, the EOS derived in this study, based on shock compression of PFH, is more robust for higher input pressures.

%Finally, we would like to emphasize the importance of selecting an appropriate range of parameters as the search space for multi-objective optimization.
%The range needs to be scientifically valid while also being sufficiently small so as to prevent the optimization run from massive computational overhead.
%This also dictates the selection of the crossover and mutation parameters, which should promote exploratory action during the offspring search, while also characterizing the focused direction in which the offspring should move every successive generation. 

\color{black}
\section{Conclusion}

Perfluorohexane shows promise to be employed as a liquid in biocompatible agents designed for ultrasound and shock wave-based therapies. 
To study the mechanisms involved in such acoustic-agent interactions, several studies focus on performing numerical research, which requires an EOS specially tuned for high-pressure conditions.
%applications using high-amplitude sound waves, such as high-intensity ultrasound and shock waves.
In this study, we have performed planar impact experiments using a single-stage mesoscale gas launcher and obtained a relationship between the shock and particle velocity for PFH.
The experiments are especially successful for low Mach number flow regimes that can be expected in biomedical applications. 
This relationship can be used as a precursor to determine other thermodynamic equations of state.
Here, we use a well-established multi-objective optimization scheme to determine the necessary constants to define the Noble-Abel Stiffened-Gas (NASG) equation, which is a convex equation of state accounting for agitation and repulsive molecular effects.
The equation is then used in hydrodynamic numerical simulations performed using ECOGEN. %to validate our EOS.
We show that we can successfully simulate the shock propagation in our impact and shock-droplet interaction experiments.
%The results were extremely satisfying, in particular for the shock-drop case, since it is a directly transferable scenario to biomedical applications. 
While the shock-droplet results presented here are for large droplets [$\approx\mathcal{O}(1~\si{\milli\meter})$] as compared to what would be used in the human body, the provided EOS enables numerical investigations of ADV at the micro- and nano-scales.
In particular, the creation and amplification of negative pressure localized around the droplet focus from a purely compressive wave reduces the possibility of potential off-target bioeffects in biomedical applications, which is why shock waves could be considered to initiate ADV.
Further simulations could help us to better understand the physical mechanisms underlying the interaction of acoustic waves with biomedical agents.
Such studies could be used to improve acoustic driving and agent formulation for biomedical applications, and in particular, accelerate the introduction of perfluorocarbon droplets undergoing ADV as theranostic agents.

\color{black}

\begin{acknowledgments}

This project was funded by the Swiss National Science Foundation (grant number 200021\textunderscore200567).
The authors acknowledge ESRF for providing resources to conduct the experiments.
The authors would also like to thank Dr. Sayaka Ichihara for helping with the shock-droplet experiments and Mr. Dengel Karakoc from the Techpool facility at ETH Z\"urich for assisting with the manufacturing process of the sabots and containers.

\end{acknowledgments}

\appendix

\section{Convergence and termination criteria for optimization runs}
\label{app:convergence}

\begin{figure*}
   \centering
    \includegraphics[width=\linewidth]{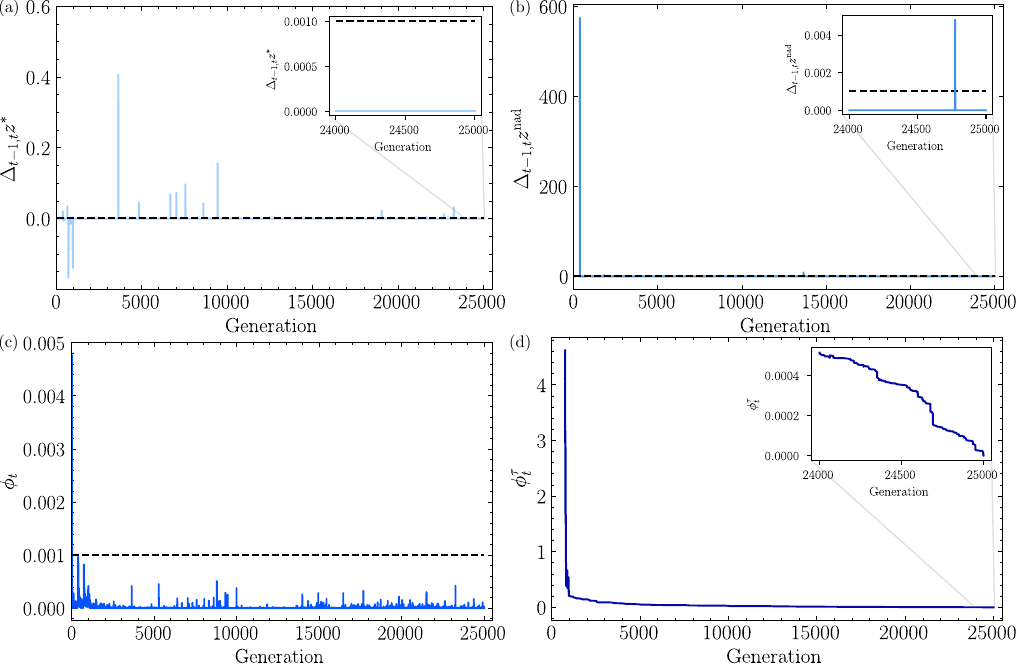}
    \caption{Evolution of the convergence criteria for determining the NASG EOS parameters for PFH. (a) Normalized change of ideal point, $\Delta_{t-1,t}z^*$, (b) Normalized change of nadir point, $\Delta_{t-1,t}z^\mathrm{nad}$, (c) Improvement of normalized IGD over successive generations, $\phi_t$. (d) Visualization IGD metric, normalized to the last generation, $\tau_\mathrm{end} = 25000$, plotted for $250\leq t_\mathrm{gen} \leq \tau_\mathrm{end}$. The quantity decreases over generations and reaches a sufficiently low value to indicate convergence of the optimization. The dotted black line in (a--c) shows the tolerance ($\epsilon=0.001$, $\chi=100$) used as the termination criteria for the optimization run.}
    \label{fig_9} 
\end{figure*}

The convergence of an optimization run is monitored using two steps: (a) convergence to extreme points and (b) convergence of solution diversity~\cite{Blank2020Pymoo:Python,Blank2020AAlgorithms}. 
The convergence to extreme points is monitored by computing the normalized change of the ideal (\emph{i.e.}, minimum) and nadir (\emph{i.e.}, maximum) points between successive generations, defined as \cite{Blank2020AAlgorithms}
\begin{eqnarray}
    \label{eq:conv_ideal}
    \Delta_{t-1,t} z^* = \max_{i=1}^M \frac{z_i^*(t-1) - z_i^*(t)}{z_i^\mathrm{nad}(t) - z_i^*(t)}\\
    \label{eq:conv_nad}
    \Delta_{t-1,t} z^\mathrm{nad} = \max_{i=1}^M \frac{z_i^\mathrm{nad}(t-1) - z_i^\mathrm{nad}(t)}{z_i^\mathrm{nad}(t) - z_i^*(t)}
\end{eqnarray}
where $z_i^*$ is the realized ideal point and $z_i^\mathrm{nad}$ is the realized nadir point for the $i$-th objective normalized at time $t$.
To monitor the diversity enhancement of the solution, a cumulative approach from the initial ($t = 0$) to the current generation ($ t=\tau$) is used. 
We first determine the normalized $i$-th objective value of the $j$-th non-dominated point at generation $t$, $\overline{P_i^{\tau}}(t)$, using the ideal and nadir points at $t=\tau$.
\begin{equation}
    \label{eq:conv_div1}
    \overline{P_i^{\tau}}(t) = \frac{P_i^j(t) - z_i^*(\tau)}{z_i^\mathrm{nad}(\tau) - z_i^*(\tau)}
\end{equation}
The improvement of the Inverted Generational Distance (IGD) metric between the normalized solution values between two successive generations is evaluated as defined below
\begin{equation}
    \label{eq:IGD_inst}
    \begin{split}
            \phi_t &= \mathrm{IGD}(\overline{P_i^{t}}(t-1),\overline{P_i^{t}}(t)) \\
            &= \frac{1}{|\overline{P_i^{t}}(t)|}\sum\limits_{i=1}^{|\overline{P_i^{t}}(t)|} \left(\min_{j=1}^{\overline{P_i^{t}}(t-1)} ||\overline{P_i^{t}}(t) - \overline{P_i^{t}}(t-1)||\right) 
    \end{split}  
\end{equation}
Eq.~\eqref{eq:IGD_inst} provides a good estimate of the broadening of the obtained solutions in successive generations. 
As a termination criterion, we ensure that the three quantities determined by Eqs.~\eqref{eq:conv_ideal}, \eqref{eq:conv_nad}, and \eqref{eq:IGD_inst} are below a certain tolerance, $\epsilon$, over a given number of transitions, by using a sliding window, $\chi$~\cite{Blank2020AAlgorithms}.
For example, for a multi-objective optimization run of PFH, we set a tolerance of $\epsilon = 0.001$ and use a window size of $\chi = 100$.
This implies that a run is found to be ``complete'' if $\Delta_{t-1,t}z^*,\;\Delta_{t-1,t}z^\mathrm{nad}\;\text{and}\;\phi_t$ vary by less than 0.1\% over the last 100 generational transitions of the run.
To visualize the convergence, it is beneficial to renormalize Eq.~\eqref{eq:IGD_inst} over the complete interval $0\leq t \leq \tau_\mathrm{end}$ and evaluate the new IGD metric as given below
\begin{equation}
    \label{eq:IGD_sum}
    \phi_t^\tau = \mathrm{IGD}(\overline{P_i^{\tau_\mathrm{end}}}(t),\overline{P_i^{\tau_\mathrm{end}}}(\tau_\mathrm{end}))
\end{equation}
While Eq.~\eqref{eq:IGD_sum} can be reevaluated every $\chi$ generations to effectively monitor convergence, this method has a high computational overhead due to the requirement of renormalizing the solution every $\chi$ generations. 
Hence, Eq.~\eqref{eq:IGD_sum} is only used as a visualization tool, while the termination is set by Eqs.~\eqref{eq:conv_ideal}, \eqref{eq:conv_nad} and \eqref{eq:IGD_inst}~\cite{Blank2020AAlgorithms}. 
Fig.~\ref{fig_9} plots $\Delta_{t-1,t}z^*$, $\Delta_{t-1,t}z^\mathrm{nad}$, $\phi_t$ and $\phi_t^{\tau}$ for one of the optimization runs. 
As depicted, all quantities are well below the set tolerance value of $\epsilon=0.001$, for at least $\chi=100$ generations, with $\phi_t^\tau$ continuously decreasing over time, indicating that the optimization has converged and can be terminated. 

\section{Uncertainty quantification of experimental data}
\label{app:error}

%Measurement of experimental quantities is usually associated with uncertainties that are either systematic or random.
Experimental errors are quantitifed in a manner described previously for such experiments by \citet{Mitchell1981ShockTantalum}. 
Briefly, an error in impact-related quantities propagates throughout the calculation due to the flyer EOS being used to determine the properties of the liquid.
Using a Monte-Carlo estimate of error propagation, for the pressure and the specific volume in Eq.~\eqref{eq:RH_definition} this corresponds to
\begin{equation}
    \begin{split}
        \label{eq:RH_err}
        \frac{\delta P}{P} &= \left[\left(\frac{\delta \rho_0}{\rho_0}\right)^2 + \left(\frac{\delta u_\mathrm{s}}{u_\mathrm{s}}\right)^2 + \left(\frac{\delta u_\mathrm{p}}{u_\mathrm{p}}\right)^2 \right]^{1/2}\\
        \frac{\delta v}{v} &= \left[\left(\frac{\delta \rho_0}{\rho_0}\right)^2 + (\eta-1)^2\left\{\left(\frac{\delta u_\mathrm{s}}{u_\mathrm{s}}\right)^2 + \left(\frac{\delta u_\mathrm{p}}{u_\mathrm{p}}\right)^2\right\}\right]^{1/2}
    \end{split}
\end{equation}
where $\eta = \rho/\rho_0$ is the compression factor, and the other terms are as defined previously. 
For the internal energy, a simple estimate can be obtained for the error by setting $P_0 = 0$, giving the error as $\delta e /e = 2 \delta u_\mathrm{p}/u_\mathrm{p}$.
For the flyer, we use $\delta u_\mathrm{p}/u_\mathrm{p} \approx \delta u_\mathrm{f}/u_\mathrm{f}$. 
This has a minor error contribution since the laser gates are well-calibrated and the sampling rate of the oscilloscope is sufficient to accurately measure the impact speed.

To determine the particle velocity in the container during impact, we include the 2\% error in the EOS of PMMA as estimated by \citet{Jordan2016ShockPolymethylmethacrylate} in their experiments. 
This introduces a systematic error in the evaluation of the flyer and the container pressures, and the uncertainty in particle velocity is taken as $\delta u_\mathrm{p} = \frac{1}{2}|u_\mathrm{p,c,max} - u_\mathrm{p,c,min}|$, where $u_\mathrm{p,c,max}$ is obtained by the intersection of $P_\mathrm{c,min}$ and $P_\mathrm{f,max}$, and $u_\mathrm{p,c,min}$ is obtained by the intersection of $P_\mathrm{c,max}$ and $P_\mathrm{f,min}$. 
The two extrema of the expected particle velocities are used to calculate the possible bounds of the release adiabats in the $P-u_\mathrm{p}$ plane, where an additional error is introduced by the term $\mathrm{d}u_\mathrm{s}/\mathrm{d}u_\mathrm{p}$ in the Riemann integral.

The largest error contribution arises from the measurement of the shock velocity from the images, $\delta u_\mathrm{s,l}/u_\mathrm{s,l}$, especially at higher speeds.
The shock velocity is estimated as $u_\mathrm{s} = \Delta x/\Delta t$, where $\Delta x$ is the distance travelled between two frames having an interval of $\Delta t$.
The error in the shock velocity estimation is taken as
\begin{equation}
    \label{eq:u_sh_err}
    \frac{\delta u_\mathrm{s,l}}{u_\mathrm{s,l}} = \left[\left(\frac{\delta (\Delta x)}{\Delta x}\right)^2 + \left(\frac{\delta (\Delta t)}{\Delta t}\right)^2 \right]^{1/2}
\end{equation}
where $\delta(\Delta x)$ is the uncertainty associated with the edge detection of the shock front and $\delta (\Delta t)$ is evaluated by taking a two-standard deviation interframe window~\cite{Possolo2015SimpleResults}.

Finally, the error on the liquid particle velocity is evaluated in a similar fashion to the container particle velocity, $\delta u_\mathrm{p,l} = \frac{1}{2}|u_\mathrm{p,l,max} - u_\mathrm{p,l,min}|$, where $u_\mathrm{p,l,max}$ is obtained by the intersection of $P_\mathrm{l,max} \propto \rho_\mathrm{PFH,max} (u_\mathrm{s,l} + \delta u_\mathrm{s,l})$ and the release adiabat evaluated from $u_\mathrm{p,c,max}$, and $u_\mathrm{p,l,min}$ is obtained by the intersection of $P_\mathrm{l,min} \propto \rho_\mathrm{PFH,min} (u_\mathrm{s,l} - \delta u_\mathrm{s,l})$ and the release isentrope evaluated from $u_\mathrm{p,c,min}$. 
Then, the errors on the thermodynamic quantities are evaluated using Eq.~\eqref{eq:RH_err}.

%\newpage

\bibliography{bibliography}% Produces the bibliography via BibTeX.

\end{document}